\documentclass[runningheads]{llncs}
\usepackage[T1]{fontenc}
\usepackage{graphicx}
\usepackage{fancyvrb}
\usepackage{fvextra}
\usepackage{todonotes}
\usepackage{array}
\usepackage{multirow}
\usepackage{listings}
\usepackage{caption}
\usepackage{graphicx}
\usepackage{booktabs}
\usepackage[export]{adjustbox}
\usepackage{enumitem}
\usepackage{makecell}
\usepackage{subscript}
\usepackage[numbers,sort&compress]{natbib}
\usepackage{wrapfig}
\usepackage{paralist}
\usepackage{dblfnote}
\usepackage{amsmath}

\usepackage[utf8]{inputenc}
\usepackage{booktabs}   
\usepackage{caption}    
\usepackage{lmodern}  
\usepackage{float}
\usepackage{enumitem}
\usepackage{multicol}
\usepackage{dblfnote}

\DFNalwaysdouble

\renewcommand{\footnotesize}{\tiny}
\newcommand{\mcrot}[4]{\multicolumn{#1}{#2}{\rlap{\rotatebox{#3}{#4}~}}} 
\begin{document}
\title{Vexed by VEX Tools: Consistency Evaluation of Container Vulnerability Scanners}
\titlerunning{Consistency Evaluation of Container Vulnerability Scanners}

\author{
Yekatierina Churakova\orcidID{0009-0004-0657-095X} \and
Mathias Ekstedt\orcidID{0000-0003-3922-9606} \and
Larissa Schmid\orcidID{0000-0002-3600-6899}
}

\authorrunning{Churakova et al.} 

\institute{
KTH Royal Institute of Technology, Stockholm, Sweden\\
\email{yekchu@kth.se, mekstedt@kth.se, lgschmid@kth.se}
}
\maketitle              

\begin{abstract}
  The Vulnerability Exploitability eXchange (VEX) format has been introduced to complement Software Bill of Materials (SBOM) with security advisories of known vulnerabilities. VEX gives an accurate understanding of vulnerabilities found in the dependencies of third-party software, which is critical for secure software development and risk analysis. 
  In this paper, we present a study that analyzes state-of-the-art VEX-generation tools (Trivy, Grype, DepScan, Scout, Snyk, OSV, Vexy) applied to containers. 
  
  Our study examines how consistently different VEX-generation tools perform. By evaluating their performance across multiple datasets, we aim to gain insight into the overall maturity of the VEX-generation tool ecosystem, beyond any single implementation. We use the Jaccard and Tversky indices to produce similarity scores of tool results for three different datasets created from container images. 
  Overall, our results show a low level of consistency among the tools, thus indicating a low level of maturity in the VEX tool space. We perform a number of experiments to explore the impact of different factors on the consistency of the results, with the difference in vulnerability databases queried showing the largest impact.

\keywords{Vulnerability Exploitability eXchange \and Software Bill of Materials \and Software Supply Chain \and Cybersecurity \and Docker Containers \and Software Vulnerability Management}

\end{abstract}
\section{Introduction}

As software products become increasingly complex, they often reuse existing tools, libraries, and dependencies. The end-to-end network of tools, libraries, dependencies, and processes involved in creating, distributing, and deploying a software product is referred to as the software supply chain \cite{what_is_ssc}. 
The visibility of libraries, modules, and packages in a supply chain can be gained through Software Bills of Materials (SBOMs).
An SBOM is a machine-readable transparent inventory of all the components, dependencies, and associated metadata that make up a software product~\cite{cisa_sbom}. 
Although an SBOM lists all the components in a software, it does not contain information about potential vulnerabilities in these components. To supplement this information, the Vulnerability Exploitability eXchange (VEX)\footnote{https://cyclonedx.org/capabilities/vex/} standard was introduced~\cite{NTIA_VEX}. VEX is an add-on to the existing SBOM specification that allows mapping the known vulnerabilities from vulnerability databases to listed components.

VEX offers a machine-readable format for tracking vulnerabilities, ensuring accurate and efficient dissemination\footnote{https://sbom.observer/academy/learn/topics/vex} and enabling rapid risk assessment and response. Its reports inform security decisions for software and dependencies, helping security teams prioritize and allocate resources by clarifying which vulnerabilities in software products, including containers, file systems, and repositories, are present based on static analysis. This can be especially useful in dynamic environments, such as containers. 

Containerization has transformed software development and deployment by offering flexibility, scalability, and portability, leading to widespread adoption across industries~\cite{container_in_software}. However, containers also introduce new security challenges~\cite{docker_sec}: They often pull third-party images from public repositories like Docker Hub\footnote{https://hub.docker.com/search?type=image}, which may contain vulnerabilities or malicious code~\cite{threats_in_supply_chains}. Moreover, their complex interdependencies also expand the attack surface and increase exploitation risks~\cite{threats_in_supply_chains}.

Given this, VEX-reports are particularly useful in the context of containers, where automated container deployment can rapidly spread compromised images and allow attackers to compromise the software supply chain. 

Although there are several VEX-generation tools, they differ in capabilities and functionality, with tools querying different vulnerability disclosure data and relying on different inputs. 
However, no insight is available into how the results of the tools compare.
As publicly available benchmarks for quality evaluation are lacking, VEX-generation tools do not provide comprehensive analysis and evaluations on public benchmarks, making it hard to assess their accuracy and complicating comparability between tools. 

Moreover, to the best of our knowledge, there is no systematic comparison of VEX-generation tools in the scientific literature. The related studies mostly evaluate some of the tools as vulnerability scanners for container images. However, the evaluation is based on the statistical distribution of vulnerabilities. The common way to interpret the results is to consider the largest number of vulnerabilities as the most precise value. This means that the overall consistency of the tools is not evaluated.
%

In this work, we systematically evaluate VEX-generation tools within the context of container images. Our analysis focuses on the consistency of these tools in identifying vulnerabilities in a container image dataset. High consistency indicates greater collective trustworthiness of the VEX-generation tools.
In detail, this paper makes the following contributions: 
\begin{inparaenum}[(i)]
\item We evaluate whether state‐of‐the‐art VEX‐generation tools identify an identical set of vulnerabilities when applied to a common reference dataset. 
We show a low level of consistency. The reported vulnerabilities differ extensively between the examined tools, and we record 69\% as the highest pairwise similarity between the tools. 
\item We investigate the factor that causes inconsistent results. According to our experiments, limiting the dataset to a single vulnerability identifier has a minor impact on similarity. We draw the same conclusion for testing tools against one specific input type.
\item We have ruled out that SBOM quality is the explanatory factor for the inconsistency of VEX-reports. 
\item We demonstrate the impact of similarity in vulnerability database integrated in VEX-generation tools on the consistency of the tools. The results show a positive correlation of 88\%, which makes this the most impactful factor. 
\end{inparaenum}
To support reproducibility, we provide our dataset and analysis data as supplementary material\footnote{https://anonymous.4open.science/r/anonymous4444/container\_hashes.txt}. 

\section{Background}

Section \ref{VEX_spec} gives an overview of the VEX-report structure and describes the connection between VEX and SBOM. Section \ref{tool_conf} describes the VEX-generation process adopted by the tools for producing the report.

\subsection{VEX} \label{VEX_spec}
The Vulnerability Exploitability eXchange (VEX)\cite{cisa_min_vex_req} is a type of security advisory aimed at conveying the exploitability of components with known vulnerabilities within the specific context of the product using them.
VEX enables software vendors and other stakeholders to communicate the exploitability status of vulnerabilities in a software product, offering clarity on which vulnerabilities present a risk and which do not. 
For our study, we focus on three of the report fields given by CISA (American Cyber Defense Agency)~\cite{cisa_min_vex_req}:
\textbf{(1) Product identifier}: Specifies the main software package that is being analyzed. This can refer to a single product, multiple products, or a product family.  
%
\textbf{(2) Vulnerability identifier}: This can be specified using a known identifier system, such as Common Vulnerabilities and Exposures (CVE) or GitHub Security Advisory (GHSA), but also using a textual description of the vulnerability. 
%
\textbf{(3) Status}: This field indicates if a specific vulnerability is active in the product. 
%
\noindent
We chose these three fields because they provide sufficient information to compare vulnerabilities from two datasets and draw a conclusion about their similarity. In our study, two different VEX-generation tools are considered consistent if they report the same identifier for the same vulnerability in the same product.

\subsection{VEX-generation Process} \label{tool_conf}

\begin{figure}[htbp]
  \centering
  \includegraphics[width=0.75\linewidth]{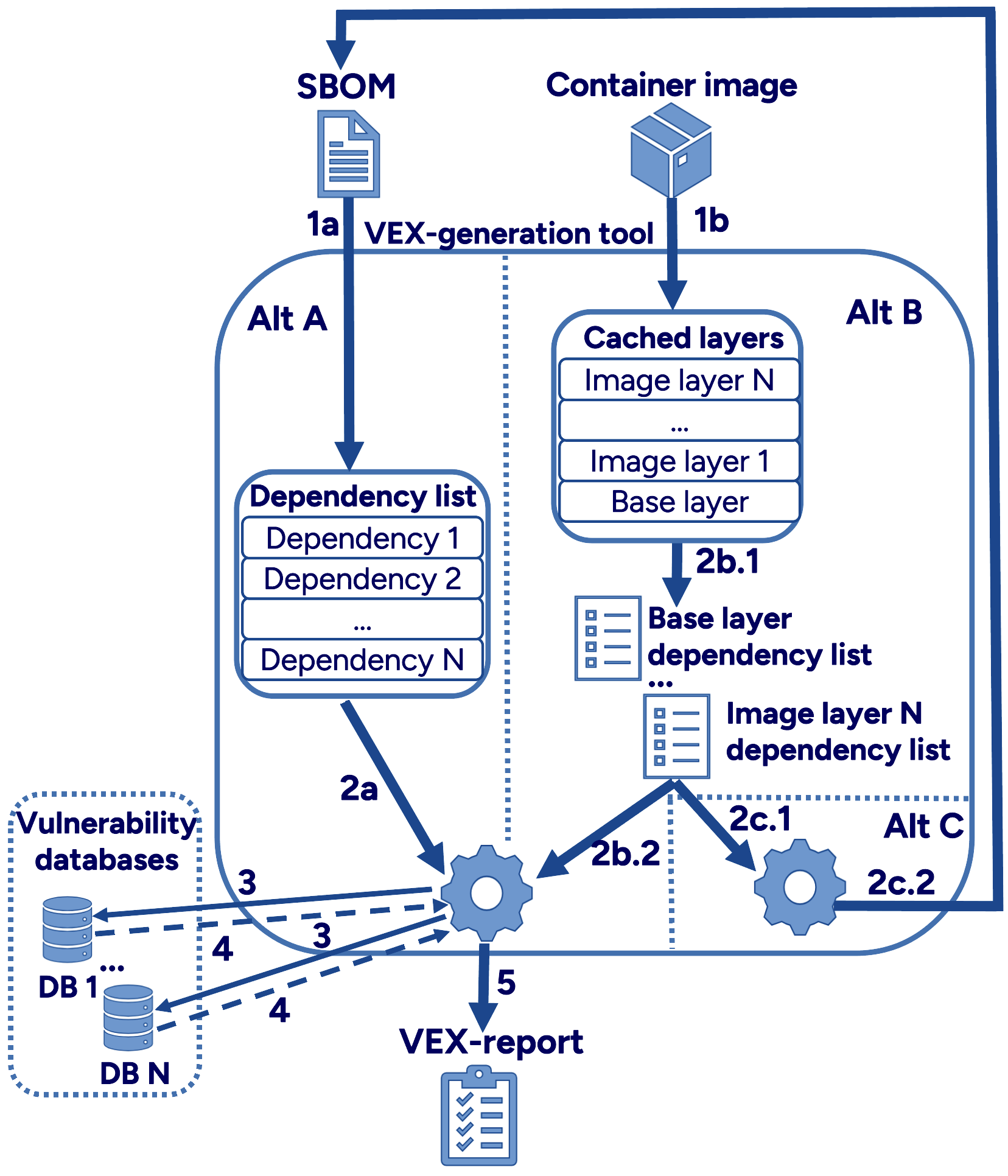} 
  \caption{Overall performance across benchmarks.}
  \label{fig:vex_gen}
\end{figure}

VEX-generation tools automatically produce VEX-reports. Figure \ref{fig:vex_gen} illustrates the three possible alternatives for generating a VEX-report for containers that differ in the input type used. 
The first one, marked as \textbf{Alt A}, is to scan the SBOM already generated for the container. 
In step \textbf{1a}, the tool reads the dependency list from the SBOM and caches the dependency list in step \textbf{2a}. The second way, represented as \textbf{Alt B} in the Figure, is direct container image scanning. In the first step, marked as \textbf{1b}, the container image layers are pulled to the tool's cache. Next, the tool analyzes the packages in each layer and creates a dependency list for each layer as step \textbf{2b.1}. Then, the whole information about dependencies is cached in step \textbf{2b.2}.

Alternatively, it is possible to produce an SBOM from the dependency lists of the different layers, constituting \textbf{Alt C}:  In step \textbf{2c.1}, the tool combines the dependency lists from the SBOM in step \textbf{2c.2}. Then, the SBOM scan by \textbf{Alt A} becomes possible.
Once the tool cached the information on the dependencies, it makes a request to the vulnerability databases to map each dependency to known vulnerabilities in step \textbf{3} and receives the answer from each vulnerability database in step \textbf{4}. After that, all the data is collected and the VEX-report is produced in the final step \textbf{5}. 

Overall, VEX-generation tools differ from each other not only in the process they follow to generate VEX-reports, but also in the number of vulnerability databases they query and identifier systems.

\section{Methodology}
The goal of our work is to evaluate the consistency of VEX-reports produced by multiple VEX-generation tools and to investigate the factors that affect this consistency. To achieve this, we select a set of VEX-generation tools and a dataset of container images to test on. This section details the selection criteria and characteristics of these artifacts.

\subsection{VEX-generation Tools Selection} \label{sec:methodology:tool-selection}

We select VEX-generation tools focused on generating VEX-reports for containers and refine the list to include only those that meet the following criteria:
%
(i) feasibility to generate lists of vulnerabilities for Docker-compatible container images or container SBOMs,
(ii) compatibility with one of the two standardized VEX formats, CycloneDX\footnote{https://cyclonedx.org/} or SPDX\footnote{https://spdx.dev/}, 
(iii) being open source, and
(iv) operating as a command-line tool. 

These last two criteria are crucial for automating our experiments and ensuring reproducibility. 
Ultimately, this process led to seven VEX-generation tools: Trivy\footnote{https://www.aquasec.com/products/trivy/}, Grype\footnote{https://github.com/anchore/grype}, DepScan\footnote{https://github.com/owasp-dep-scan/dep-scan}, OSV-scanner\footnote{https://osv.dev/\#use-vulnerability-scanner}, Docker Scout utility\footnote{https://docs.docker.com/scout/}, Vexy\footnote{https://github.com/madpah/vexy}, and Snyk\footnote{https://snyk.io/lp/sast-tools}. We use the most recent stable releases of these tools at the time of the study and provide their exact versions as part of the supplementary material.

As mentioned in Section \ref{tool_conf}, there are three different alternatives to generate VEX-reports. Not all tools can be configured to follow all possible ways (cf. Table \ref{tab:tools}): Vexy can only scan existing SBOMs but cannot scan container images or produce SBOMs. In contrast, Docker Scout cannot scan external SBOMs, but can scan container images and produce SBOMs. Trivy, Grype and DepScan can be configured for all three alternatives. We mark Snyk with a --(+) sign because it offers SBOM production and scanning features exclusively in enterprise subscriptions at the time of the study, but not in the open-source command line version we used. OSV-scanner is marked with a --(+) as it provides container image scanning only for Debian-based images.

\begin{table}
\centering
\begin{adjustbox}{width=.6\linewidth}
  \centering
  \begin{tabular}{lccccccc}
    Capability       & {Trivy} & {Grype} & {DepScan} & {OSV}  & {Vexy} & {Scout} & {Snyk}\\
    \midrule
    Scan SBOMs      & +     & +     & +      & +    & +    & --     & --(+)  \\
    Produce SBOMs   & +     & +     & +      & --    & --    & +     & --(+)  \\
    Scan image      & +     & +     & +      & --(+) & --    & +     & +     \\
    \bottomrule
  \end{tabular}
\end{adjustbox}
  \caption{Tool capabilities comparison. A "+" indicates a supported feature, and a "--" indicates an unsupported feature.}
  \label{tab:tools}
\end{table}

Thus, even though we collected seven tools for this study, we have a larger number of possible tool configurations, 25 in total. Our experiments include all possible combinations of tools that can generate an SBOM (four tools) and those that can scan it (five tools), resulting in 20 combinations. In addition, we conduct five experiments using tools that can scan containers directly. 

As observed, no single input type works for all tools. This leads us to establish a baseline for our study. For tools capable of scanning container images, we select that option. This includes all tools except OSV-Scanner and Vexy, for which we use SBOMs generated by Docker Scout as the baseline. We choose Docker Scout as the SBOM generator because it is source-native to the containers in our dataset.
For consistency, we use the same SBOM versions (CycloneDX 1.4 and SPDX 2.3) in all our tests to minimize the impact of variations in dependency lists~\cite{sbom_vuln}.

\subsection{Dataset Creation}

We use Docker container images for our dataset. They bundle software and dependencies consistently, creating controlled environments with rich codebases and dependency lists. 

We form a dataset comprising 48 Docker containers sourced from Docker Hub\footnote{https://hub.docker.com/} and provide the list of specific container hashes used as part of our supplementary material to facilitate the reproducibility of the results. 
For the experiments in our study, we consider the nature of the examined containers to be out of scope, as our goal is to evaluate whether the tools find the same vulnerabilities for the dataset. Because of this, we select 32 containers randomly (the 'Random Set'). In addition, to get an understanding of the edge cases, we extend the dataset with what we expect to be a high and a low number of vulnerabilities. We include eight containers identified as completely free of vulnerabilities according to Docker Hub registry (referred to as 'Non-Vulnerable Set'), 
and an additional eight containers harboring the highest count of vulnerabilities according to the same source (the 'Vulnerable Set'). We note that the vulnerability count here is not necessarily correct. However, we trust the data enough to construct two datasets that have a high and low number of vulnerabilities, respectively.

\section{Analysis and Results} 
\label{analysis_and_results}
To evaluate the consistency in the list of vulnerabilities reported by different VEX-generation tools, we address the following research question:\\ 
\\
\renewcommand{\labelenumi}{\textbf{RQ\arabic{enumi}:}}
\begin{inparaenum}
    \item What are the pairwise similarity scores of the tools' results?\\
    \item Is there a subset of vulnerabilities detected by several tools?\\
    \item What is the impact of the input SBOM on similarity scores?\\
    \item Do the tools find the same set of vulnerabilities with the 'affected' status?\\
    \item What is the impact of vulnerability identifiers on the similarity scores of the tools' results?\\
    \item What is the correlation between vulnerability databases queried by the tools and their pairwise similarity scores?\\
\end{inparaenum}

The results of RQ1 (Section~\ref{Overall_consistency}) 
indicate that the tools' results show a low level of consistency. To identify the cause for this, we conduct a series of experiments: First, we focus on non-binary similarity (RQ2, Section \ref{analysis:non-binary-similarity}), investigating if there is a set of vulnerabilities that all tools identify. Next, we examine whether the tools demonstrate higher similarity in their results when provided the same input SBOMs (RQ3, Section \ref{analysis:input-dep}). 
Section \ref{analysis:affected-vuln} examines the similarity of vulnerabilities found with the 'affected' status (RQ4). 
We then test the impact of different vulnerability identifiers used by different tools on their similarity results (RQ5) in Section \ref{analysis:impact-identifiers}. 
Finally, Section \ref{analysis:db} looks at the impact of different vulnerability databases used by the tools on their similarity scores (RQ6). 

\subsection{Overall Consistency} \label{Overall_consistency}
Table ~\ref{tab:vuln_compl_set} shows the results of applying the VEX-generation tools to the different subsets of container images. The agreement between the tools on the number of vulnerabilities is generally low, with the number of detected vulnerabilities ranging from 191 to 18680 on the complete dataset. However, Trivy and Grype report similar counts, suggesting some alignment. However, numbers alone do not confirm whether they identify the same issues. For example, if all of Vexy's 191 findings are contained in DepScan's 18680 findings, their consistency would be stronger than between Trivy and Grype, even if Trivy and Grype report similar totals but no overlapping vulnerabilities. 
The number of vulnerabilities per container in the random subset ranges from 0.9 to 272.9. In the vulnerable subset, the number ranges from 18.5 to 609, and is on average higher than in the random subset. For the non-vulnerable subset, even though vulnerabilities were found, the number of vulnerabilities per container ranges from 0.125 to 162.25 and is lower than in the random subset.
All tools have identified vulnerabilities in the container images of the non-vulnerable subset. Furthermore, scanning the images with Scout detects 14 vulnerabilities, compared to the initial zero listed on Docker Hub.


To explore consistency, we use the Jaccard index, also known as the Jaccard similarity coefficient~\cite{jaccard_index}. This statistical measure is used to evaluate the similarity and diversity of the sample sets. It is calculated by dividing the size of the intersection of two sets by the size of the union of the sets as follows: \( J(A,B) = \tfrac{|A \cap B|}{|A \cup B|} \) (1),
where A and B are two sets. The Jaccard index ranges from 0 to 1, where 0 indicates no similarity between the sets and 1 indicates identical sets. The index provides a clear indication, and is also robust to variations in the size of the sets, focusing solely on the relative overlap rather than the absolute values. 

We calculate the Jaccard index between a pair of tools by calculating the similarity of the vulnerabilities identified by them. We consider a vulnerability to be the same if the vulnerability identifier and vulnerable component identifier from the tools' VEX-reports match between vulnerabilities. 
\ref{tab:bin_jaccard} presents the results of this analysis: Overall, we observe low consistency between tools, with many cases of very low overlap with only a few percent, and even cases of completely disjoint sets. Only Grype and Trivy demonstrate somewhat consistent results with a Jaccard index of 0.694.

\begin{table}[t]
\centering

\begin{minipage}[t]{.48\linewidth}
\centering\footnotesize
\begin{adjustbox}{width=\linewidth}
\begin{tabular}{lrrrrrrr}
  Subset & \mcrot{1}{l}{70}{Trivy} & \mcrot{1}{l}{70}{Grype} & \mcrot{1}{l}{70}{DepScan} & \mcrot{1}{l}{70}{Scout} & \mcrot{1}{l}{70}{Snyk} & \mcrot{1}{l}{70}{OSV} & \mcrot{1}{l}{70}{Vexy}\\
  \midrule
  Complete   & 12288 & 12878 & 18680 & 3036 & 15882 & 1426 & 191 \\
  Random     &  8733 &  8576 & 13097 & 1622 & 12015 &  439 &  29 \\
  Vulnerable &  2504 &  2883 &  4872 & 1367 &  3365 &  940 & 148 \\
  Non-Vuln.  &  1052 &  1298 &   338 &   14 &    80 &   35 &   1 \\
  \bottomrule
\end{tabular}
\end{adjustbox}
\captionof{table}{Number of vulnerabilities found by different tools per set.}
\label{tab:vuln_compl_set}
\end{minipage}\hfill
\begin{minipage}[t]{.48\linewidth}
\centering\footnotesize
\begin{adjustbox}{width=\linewidth}
\begin{tabular}{llllllll}
  & \mcrot{1}{l}{70}{Trivy} & \mcrot{1}{l}{70}{Grype} & \mcrot{1}{l}{70}{DepScan} & \mcrot{1}{l}{70}{Scout} & \mcrot{1}{l}{70}{Snyk} & \mcrot{1}{l}{70}{OSV} & \mcrot{1}{l}{70}{Vexy} \\
  \midrule
  Trivy   & 1     &        &       &       &       &       &     \\
  Grype   & 0.694 & 1      &       &       &       &       &     \\
  DepScan & 0.160 & 0.155  & 1     &       &       &       &     \\
  Scout   & 0.329 & 0.304  & 0.062 & 1     &       &       &     \\
  Snyk    & 0.379 & 0.355  & 0.118 & 0.332 & 1     &       &     \\
  OSV     & 0.059 & 0.004  & 0.010 & 0.129 & 0.003 & 1     &     \\
  Vexy    & 0.018 & 0      & 0.003 & 0.041 & 0     & 0.095 & 1   \\
  \bottomrule
\end{tabular}
\end{adjustbox}
\captionof{table}{Jaccard index of tool pairs for the complete dataset.}
\label{tab:bin_jaccard}
\end{minipage}
\vspace{-30pt}
\end{table}

\subsection{Non-binary Similarity} \label{analysis:non-binary-similarity}
Next, we approach consistency from the opposite angle and investigate if there are common results that most tools agree on. For this analysis, we use the Tversky index~\cite{tversky_index}, which generalizes the Jaccard index to multiple sets: 

\( J(A_1,\ldots,A_n) = \frac{\lvert \bigcap_{i=1}^{n} A_i \rvert}{\lvert \bigcup_{i=1}^{n} A_i \rvert} \) (2).

\noindent
The Tversky index calculates the value of overlapping between several sets (i.e., more than 2), providing a range from 0 to 1, where 0 indicates no similarities and 1 indicates that the sets are similar. 

Our results in Section \ref{Overall_consistency} show that Vexy has little overlap with the other tools, so we would not learn anything new by only calculating the Tversky index for the seven tools jointly. Instead, we form unions of all combinations of five and six tools as references for the calculations. In Table \ref{tab:agreement5}, we present Tversky indices for all combinations of agreement sets for five tools that are not equal to zero. We excluded from the table the list of combinations (-Trivy \& Grype, -Trivy \& DepScan, -Trivy \& Scout, -Trivy \& Snyk, -Trivy \& OSV, -Grype \& DepScan, -Grype \& Scout, -Grype \& OSV, -DepScan \& Scout, -DepScan \& OSV, -DepScan \& Snyk, -Snyk \& Scout, -Snyk \& OSV, -Scout \& OSV), as the similarity scores are equal to zero. This fact further illustrates how little the overall overlap in results is. The highest number we get is 3.4\%, which is how much Grype, Trivy, Depscan, Snyk, and Docker Scout agree on. This metric clearly shows how different the vulnerability lists in the reports are.

Table \ref{tab:agreement6} illustrates the agreement sets for each combination of six tools. We see that Vexy is the outlier in this respect, as the other six tools have several vulnerabilities in overlapping datasets, while Vexy does not report any similarities with other tools. 

\begin{table}[t]
\centering

\begin{minipage}[t]{.55\linewidth}
\centering\footnotesize
\begin{adjustbox}{width=\linewidth}
\begin{tabular}{lrrrrrrrr}
  Subset & \mcrot{1}{l}{70}{-Vexy \& Grype} & \mcrot{1}{l}{70}{-Vexy \& Trivy} &
  \mcrot{1}{l}{70}{-Vexy \& DepScan} & \mcrot{1}{l}{70}{-Vexy \& Snyk} &
  \mcrot{1}{l}{70}{-Vexy \& OSV} & \mcrot{1}{l}{70}{-Vexy \& Scout} &
  \mcrot{1}{l}{70}{-Snyk \& Grype}\\
  \midrule
  Complete   & 0.0003 & 0.0003 &        & 0.0007 & 0.034  & 0.0003 & 0.002 \\
  Random     & 0.0004 & 0.0005 & 0.001  & 0.0007 & 0.044  & 0.0004 & 0.0006 \\
  Vulnerable & 0.0001 & 0.0001 & 0.0003 & 0.0009 & 0.022  & 0.0001 & 0.005 \\
  Non-vuln.  & 0      & 0      & 0      & 0      & 0.0009 & 0      & 0 \\
  \bottomrule
\end{tabular}
\end{adjustbox}
\captionof{table}{Tversky index for groups of 5 tools.}
\label{tab:agreement5}
\end{minipage}\hfill
\begin{minipage}[t]{.4\linewidth}
\centering\footnotesize
\begin{adjustbox}{width=\linewidth}
\begin{tabular}{lrrrrrrr}
  Subset & \mcrot{1}{l}{70}{All-Trivy} & \mcrot{1}{l}{70}{All-Grype} & \mcrot{1}{l}{70}{All-DepScan} & \mcrot{1}{l}{70}{All-OSV} & \mcrot{1}{l}{70}{All-Vexy} & \mcrot{1}{l}{70}{All-Scout} & \mcrot{1}{l}{70}{All-Snyk} \\
  \midrule
  Complete   & 0 & 0 & 0 & 0 & 0.0003 & 0 & 0 \\
  Random     & 0 & 0 & 0 & 0 & 0.0004 & 0 & 0 \\
  Vulnerable & 0 & 0 & 0 & 0 & 0.00015 & 0 & 0 \\
  Non-vuln.  & 0 & 0 & 0 & 0 & 0 & 0 & 0 \\
  \bottomrule
\end{tabular}
\end{adjustbox}
\captionof{table}{Tversky index for groups of 6 tools.}
\label{tab:agreement6}
\end{minipage}
\vspace{-25pt}
\end{table}

\subsection{Input Dependency} \label{analysis:input-dep}
As shown in previous works \cite{SBOM_python2,SBOM_challenges,Balliu_2023}, generating high-quality SBOMs is challenging. In Section~\ref{sec:methodology:tool-selection}, we note that the baseline tool configurations differ in the input types they process, which could contribute to the observed inconsistencies. To examine this, we perform an experiment that assess the outcomes of the three tools, Grype, Trivy, and DepScan, which accept both container images and SBOMs as input.
We execute the tools using three types of input on the complete set of vulnerabilities:
\renewcommand{\labelenumi}{\textbf{\arabic{enumi}:}}
\begin{inparaenum}
\item Direct image scanning mode
\item Scanning of SBOMs generated by the same tool (referred to as "native" SBOMs)
\item Scanning of SBOMs produced by Docker Scout
\end{inparaenum}
However, testing DepScan with Docker Scout SBOM input was not possible due to format incompatibility: DepScan supports CycloneDX, whereas Docker Scout produces SBOMs in SPDX format, and converting SBOMs between the two formats is still a challenging task. Thus, this experiment includes eight possible scenarios.

To calculate similarity scores, we use the Jaccard index and receive the results presented in Table \ref{tab:different_inputs}. Overall, we observed low similarity even when using the same tool in direct image scanning mode and scanning its native SBOM. 
We observe the highest similarity score (86.7\%) for DepScan when comparing direct image scanning to its native SBOM.

\begin{table}
\vspace*{-20pt}
\centering
  \begin{adjustbox}{width=.8\linewidth}
  {\LARGE
  \begin{tabular}{l
  >{\raggedleft\arraybackslash}b{1.5cm}
  >{\raggedleft\arraybackslash}b{1.5cm}
  >{\raggedleft\arraybackslash}b{1.5cm}
  >{\raggedleft\arraybackslash}b{1.5cm}
  >{\raggedleft\arraybackslash}b{1.5cm}
  >{\raggedleft\arraybackslash}b{1.5cm}
  >{\raggedleft\arraybackslash}b{1.5cm}
  >{\raggedleft\arraybackslash}b{1.5cm}
  >{\raggedleft\arraybackslash}b{1.5cm}}

    & \mcrot{1}{l}{60}{Trivy} & \mcrot{1}{l}{60}{Grype} & \mcrot{1}{l}{60}{Depscan} & \mcrot{1}{l}{60}{Trivy\textsubscript{Scout}} & \mcrot{1}{l}{60}{Grype\textsubscript{Scout}} & \mcrot{1}{l}{60}{Depscan\textsubscript{Scout}} & \mcrot{1}{l}{60}{Trivy\textsubscript{Trivy}} & \mcrot{1}{l}{60}{Grype\textsubscript{Grype}} & \mcrot{1}{l}{60}{Depscan\textsubscript{Depscan}} \\
    \midrule
    Trivy\textsubscript{Scout} & { 0.075 } & {0.005} & {0.008} & 1 &&&&& \\
    Grype\textsubscript{Scout} & { 0.003 } & {0.076} & {0.0005} & {0.026} & 1 &&&& \\
    Depscan\textsubscript{Scout} & { 0.04 } & {0.047} & {0.115} & {0.014} & {0.01} & 1 &&& \\
    Tryivy\textsubscript{Trivy} & { 0.551 } & {0.471} & {0.286} & {0.058} & {0.003} & {0.063}  & 1 && \\
    Grype\textsubscript{Grype} & { 0.436 } & {0.589} & {0.275} & {0.004} & {0.053} & {0.067} & {0.083} & 1 & \\
    Depscan\textsubscript{Depscan} & 0.164 & 0.159 & 0.867 & 0.009 & 0.0005 & 0.128 & 0.3 & 0.287 & 1 \\
    \bottomrule
  \end{tabular} }
  \end{adjustbox}
    \caption{Jaccard index for three tools with different inputs for the complete dataset, with subscript indicating the tool used for creating the input SBOM, tools without subscript are used in a container scanning mode}
    \label{tab:different_inputs}
\end{table}

For SBOMs to be the explaining factor for the low consistency in the tools' results, scanning the same SBOM with different tools should lead to highly similar or even identical results. Therefore, we execute all tools accepting external SBOMs on the same SBOM and show the results in Tables \ref{tab:trivy-sbom-tool},  ~\ref{tab:grype-sbom-tool}, ~\ref{tab:depscan-sbom-tool}, and ~\ref{tab:scout-sbom-tool}, respectively, again with overall low consistencies not exceeding 70\% for the best pair performance between Grype and Trivy. 
Our results demonstrate that SBOM quality is not the explanatory factor for the inconsistency of VEX-reports and that the challenge of inconsistent VEX-reports is broader than that. 

\begin{table}[t]
\centering

\begin{minipage}[t]{.48\linewidth}
\centering\footnotesize
\begin{tabular}{lrrrrr}
  & {Trivy} & {Grype} & {DepScan} & {OSV} & {Vexy} \\
  \midrule
  Trivy & 1 & & & & \\
  Grype & 0.101 & 1 & & & \\
  DepScan & 0.087 & 0.076 & 1 & & \\
  OSV & 0.059 & 0.012 & 0.034 & 1 & \\
  Vexy & 0.003 & 0 & 0.0004 & 0.035 & 1 \\
  \bottomrule
\end{tabular}
\captionof{table}{Jaccard index for different tools with Trivy SBOM.}
\label{tab:trivy-sbom-tool}
\end{minipage}\hfill
\begin{minipage}[t]{.48\linewidth}
\centering\footnotesize
\begin{tabular}{lrrrrr}
  & {Trivy} & {Grype} & {DepScan} & {OSV} & {Vexy} \\
  \midrule
  Trivy & 1 & & & & \\
  Grype & 0.097 & 1 & & & \\
  DepScan & 0.137 & 0.091 & 1 & & \\
  OSV & 0.278 & 0.004 & 0.062 & 1 & \\
  Vexy & 0.065 & 0 & 0.013 & 0.102 & 1 \\
  \bottomrule
\end{tabular}
\captionof{table}{Jaccard index for different tools with Grype SBOM.}
\label{tab:grype-sbom-tool}
\end{minipage}


\begin{minipage}[t]{.48\linewidth}
\centering\footnotesize
\begin{tabular}{lrrrrr}
  & {Trivy} & {Grype} & {DepScan} & {OSV} & {Vexy} \\
  \midrule
  Trivy & 1 & & & & \\
  Grype & 0.044 & 1 & & & \\
  DepScan & 0.008 & 0.0005 & 1 & & \\
  OSV & 0.401 & 0.022 & 0.011 & 1 & \\
  Vexy & 0 & 0 & 0 & 0 & 1 \\
  \bottomrule
\end{tabular}
\captionof{table}{Jaccard index for different tools with DepScan SBOM.}
\label{tab:depscan-sbom-tool}
\end{minipage}\hfill
\begin{minipage}[t]{.48\linewidth}
\centering\footnotesize
\begin{tabular}{lrrrrr}
  & {Trivy} & {Grype} & {DepScan} & {OSV} & {Vexy} \\
  \midrule
  Trivy & 1 & & & & \\
  Grype & 0.026 & 1 & & & \\
  DepScan & 0.014 & 0.001 & 1 & & \\
  OSV & 0.439 & 0.015 & 0.005 & 1 & \\
  Vexy & 0.003 & 0 & 0.003 & 0.017 & 1 \\
  \bottomrule
\end{tabular}
\captionof{table}{Jaccard index for different tools with Scout SBOM.}
\label{tab:scout-sbom-tool}
\end{minipage}
\vspace{-20pt}
\end{table}

\subsection{Affected Vulnerabilities} \label{analysis:affected-vuln}
\label{affected_vulns}

For this experiment, we assume that the tools should be more consistent in identifying more relevant and critical vulnerabilities with the 'affected' status, which indicates active, unpatched vulnerabilities. To perform the experiments on estimating similarity scores for the vulnerabilities with status 'affected', we filter the VEX-reports of the tools to only contain vulnerabilities marked as 'affected' in the 'status' field of the VEX-report. This is not possible for OSV and Vexy, as their reports do not contain a specific field for the status. Given that status information is part of VEX-specification, these two tools violate output requirements, making their reports less helpful in vulnerability prioritization.
However, the results in Table~\ref{tab:affected} still illustrate a high diversity in the number of vulnerabilities with 'affected' status found by each tool. For the non-vulnerable subset, DepScan and Snyk detect zero vulnerabilities with 'affected' status, which partially corresponds to the information from Docker Hub.

\begin{table}
\centering
  \begin{tabular}{lrrrrr}
    {Subset}&{Trivy}&{Grype}&{DepScan}&{Scout}&{Snyk}\\
    \midrule
    Random&7767&5787&2151&1266&1671\\
    Vulnerable&1682&1241&611&380&793\\
    Non-vulnerable&42&112&0&2&0\\
  \bottomrule
\end{tabular}
  \caption{Number of vulnerabilities with 'affected' status.}
  \label{tab:affected}
  \vspace{-30pt}
\end{table}

\subsection{Impact of Vulnerability Identifiers} \label{analysis:impact-identifiers}

In our effort to identify factors contributing to the inconsistencies recorded, we now examine the diversity of vulnerability identifiers and references used by the tools. Within our complete set of vulnerabilities, we find several types of vulnerability identifiers: CVE, GHSA, NSWG (Node.js Security Working Group), BIT (Bitnami Security Advisories), DSA (Debian Security Advisory), NPM (Node Package Manager), and TEMP. 
First, we acknowledge that the TEMP identifier is a placeholder commonly used to track vulnerabilities, threat actors, or attack patterns that have not yet been formally categorized or assigned a known identifier. It is typically used until an official identifier, such as a CVE, is assigned by the relevant authority. As a result, vulnerabilities tagged with TEMP could significantly impact the consistency of VEX-reports. To mitigate this effect, we calculate the Jaccard index for our dataset after excluding vulnerabilities with TEMP identifiers. The number of filtered entities differs from 0 to 126 TEMP identifiers across the dataset. The results in Table~\ref{tab:bin_jaccard-temp} show some consistency growth for some pairs of tools, with the highest increase between Trivy and Snyk (+2.10\%). 


Next, we also take a bottom-up approach: In our previous experiments, different identifier systems are inherently counted as mismatches. As a result, comparing tools that rely on different identifier systems is not meaningful. Therefore, we calculate similarity scores for the reported vulnerabilities with only the most common identifiers CVE and GHSA. CVE presents 98\% of the vulnerabilities found by the tools, as shown in Table ~\ref{tab:vuln_compl_set_cve}. 
This means that we do not consider the 2\% of vulnerabilities using other identifiers for this experiment. 



\begin{table}[H]
\centering\small
\setlength{\tabcolsep}{1pt}
\begin{tabular}{lrrrrrrr}
  & \mcrot{1}{l}{70}{Trivy} & \mcrot{1}{l}{70}{Grype} & \mcrot{1}{l}{70}{DepScan} & \mcrot{1}{l}{70}{Scout} & \mcrot{1}{l}{70}{Snyk} & \mcrot{1}{l}{70}{OSV} & \mcrot{1}{l}{70}{Vexy}\\
  \midrule
  Trivy   & –      & +1.6\% & +0.1\% & +1.1\% & +2.1\% & +0.1\% & +0.0\% \\
  Grype   & 0.71   & –      & +0.0\% & +0.0\% & +0.0\% & +0.0\% & +0.0\% \\
  DepScan & 0.161  & 0.155  & –      & +0.0\% & +0.2\% & +0.0\% & +0.1\% \\
  Scout   & 0.34   & 0.304  & 0.062  & –      & +0.0\% & +0.0\% & +0.0\% \\
  Snyk    & 0.40   & 0.355  & 0.12   & 0.332  & –      & +0.0\% & +0.0\% \\
  OSV     & 0.06   & 0.004  & 0.010  & 0.129  & 0.003  & –      & +0.0\% \\
  Vexy    & 0.018  & 0      & 0.004  & 0.041  & 0      & 0.095  & –      \\
  \bottomrule
\end{tabular}
\captionof{table}{Jaccard index for tool pairs for the complete dataset excluding TEMP entries — lower triangle. Upper triangle: change in \% vs. Table \ref{tab:bin_jaccard}.}
\label{tab:bin_jaccard-temp}
\end{table}
\vspace{-30pt}
\begin{table}[H]
\centering\small
\setlength{\tabcolsep}{1pt}
\begin{tabular}{lrrrrrrr}
  & \mcrot{1}{l}{70}{Trivy} & \mcrot{1}{l}{70}{Grype} & \mcrot{1}{l}{70}{DepScan} & \mcrot{1}{l}{70}{Scout} & \mcrot{1}{l}{70}{Snyk} & \mcrot{1}{l}{70}{OSV} & \mcrot{1}{l}{70}{Vexy}\\
  \midrule
  Complete set  & 98.6 & 93.2 & 97.9 & 96.8 & 100.0 & 52.2 & 100.0 \\
  Random set    & 98.7 & 99.0 & 99.9 & 98.6 & 100.0 & 34.6 & 100.0 \\
  Vuln set      & 97.7 & 72.6 & 97.3 & 94.5 & 100.0 & 61.2 & 100.0 \\
  Non-vuln set  & 100.0& 100.0& 100.0& 100.0& 100.0 & 54.3 & 100.0 \\
  \bottomrule
\end{tabular}
\captionof{table}{Percentage of CVE entries among all vulnerabilities (values in \%).}
\label{tab:vuln_compl_set_cve}
\end{table}

%
Table~\ref{tab:bin_jaccard_cve} shows the resulting Jaccard indexes between tools for CVE identifiers and Table~\ref{tab:bin_jaccard_ghsa} for GHSA identifiers, respectively.

Compared to Table~\ref{tab:bin_jaccard}, some tool pairs show higher similarity for single-identifier sets, while others show a decrease in similarity. Although our experiments show that the presence of different vulnerability identifiers affects the similarity scores of the reports, the impact is not consistent.

\begin{table}[H]
\centering\small
\begin{tabular}{lrrrrrrr}
  & \mcrot{1}{l}{70}{Trivy} & \mcrot{1}{l}{70}{Grype} & \mcrot{1}{l}{70}{DepScan} & \mcrot{1}{l}{70}{Scout} & \mcrot{1}{l}{70}{Snyk} & \mcrot{1}{l}{70}{OSV} & \mcrot{1}{l}{70}{Vexy}\\
  \midrule
  Trivy   & –      & +6.6\% & +0.3\%  & +0.1\%  & +0.8\%  & +0.3\%  & –0.3\%  \\
  Grype   & 0.76   & –      & +0.7\%  & +0.0\%  & +0.0\%  & +0.0\%  & +0.0\%  \\
  DepScan & 0.163  & 0.162  & –       & +0.0\%  & +0.0\%  & –0.3\%  & +0.1\%  \\
  Scout   & 0.33   & 0.304  & 0.062   & –       & +0.8\%  & –0.3\%  & –1.1\%  \\
  Snyk    & 0.387  & 0.355  & 0.118   & 0.34    & –       & +0.0\%  & +0.0\%  \\
  OSV     & 0.062  & 0.004  & 0.007   & 0.126   & 0.003   & –       & +8.5\%  \\
  Vexy    & 0.015  & 0      & 0.004   & 0.03    & 0       & 0.18    & –       \\
  \bottomrule
\end{tabular}
\captionof{table}{Increase of Jaccard index incl. CVE-only identifiers (upper triangle, \% vs. Table \ref{tab:bin_jaccard}) and exact Jaccard indices (lower triangle).}
\label{tab:bin_jaccard_cve}
\end{table}
\vspace{-20pt}
\begin{table}[H]
\centering\small
\begin{tabular}{lrrrrrrr}
  & \mcrot{1}{l}{70}{Trivy} & \mcrot{1}{l}{70}{Grype} & \mcrot{1}{l}{70}{DepScan} & \mcrot{1}{l}{70}{Scout} & \mcrot{1}{l}{70}{Snyk} & \mcrot{1}{l}{70}{OSV} & \mcrot{1}{l}{70}{Vexy}\\
  \midrule
  Trivy   & –      & -65.4\% & +19.0\% & +37.1\% & -37.9\% & +48.1\% & -1.8\% \\
  Grype   & 0.04   & –       & -13.5\% & -25.4\% & -35.5\% & +3.6\%  & +0.0\% \\
  DepScan & 0.35   & 0.02    & –       & +22.8\% & -11.8\% & +12.0\% & -0.3\% \\
  Scout   & 0.7    & 0.05    & 0.29    & –       & -33.2\% & +0.1\%  & -4.1\% \\
  Snyk    & 0      & 0       & 0       & 0       & –       & -0.3\%  & +0.0\% \\
  OSV     & 0.54   & 0.04    & 0.13    & 0.13    & 0       & –       & -9.5\% \\
  Vexy    & 0      & 0       & 0       & 0       & 0       & 0       & –      \\
  \bottomrule
\end{tabular}
\captionof{table}{Increase of Jaccard index incl. GHSA-only identifiers (upper triangle, \% vs. Table \ref{tab:bin_jaccard}) and exact Jaccard indices (lower triangle).}
\label{tab:bin_jaccard_ghsa}
\end{table}

\subsection{Vulnerability Databases Effect on Similarity} \label{analysis:db}

Given the variety of vulnerability identifiers in the reports, we hypothesize that references to different databases, such as NVD (U.S. government vulnerability repository), GHSA (the acronym refers both to vulnerability database and to vulnerability identifiers assigned by this database), and NPM (the NPM vulnerability database, which tracks security issues in packages), influence the similarity scores between VEX-reports.

VEX-generation tools identify vulnerabilities in dependencies of a container image by making requests to vulnerability databases (cf. Section \ref{tool_conf}). If they refer to different databases, inconsistencies may arise, even if the vulnerability identifier system is similar. This happens due to the naming problem and different vulnerability identifiers. The naming problem arises because different vulnerability databases and security tools may use varying names, formats, or identifier systems for the same vulnerability, leading to inconsistencies in tracking and reporting. For example, the GHSA-q2x7-8rv6-6q7h vulnerability (Jinja sandbox breakout)\footnote{https://osv.dev/vulnerability/GHSA-q2x7-8rv6-6q7h} has the alias CVE-2024-56326\footnote{https://nvd.nist.gov/vuln/detail/CVE-2024-56326} in NVD and SNYK-CHAINGUARDLATEST-LOCALSTACK-8679223 in Snyk database\footnote{https://security.snyk.io/vuln/SNYK-CHAINGUARDLATEST-LOCALSTACK-8679223}, so it can be mapped in the report in three possible ways. Although the connection between the GHSA and CVE identifiers can be retrieved through the OSV database, a direct mapping between the SNYK identifiers and CVE is missing. This discrepancy makes it difficult to correlate vulnerabilities across multiple databases, even when they refer to the same issue. We have investigated the documentation of the tools and retrieved the vulnerability databases they refer to (cf. Table~\ref{tab:number_of_databases}). Then, we have calculated the similarity scores for the tools based on their vulnerability database references, illustrated in Table~\ref{tab:database_similarity}. The similarity for this test is based on the vulnerability database name only. Thus, we do not consider other parameters, such as database volume, number of vulnerabilities registered, etc. We see quite a low overlap overall, with a few pairs just below 30\% at best. To identify if database similarity could be an explanatory factor for the inconsistency, we calculated the Pearson correlation\footnote{Pearson correlation is a standard statistical measure that quantifies the linear relationship between two variables on a scale from $1$, perfect positive linear relationship, and $-1$, perfect negative linear relationship. For our calculations, we use a Pandas module: https://pandas.pydata.org/docs/reference/api/ pandas.DataFrame.corr.html} between the two parameters: Similarity in databases to which the tool refers (see Table~\ref{tab:database_similarity}), and similarity in reports (see Table~\ref{tab:bin_jaccard-temp}). This results in a Pearson coefficient of 0.88, indicating a strong positive correlation between the two variables analyzed.

\begin{table}[t]
\centering

\begin{minipage}[t]{.47\linewidth}
\centering\tiny
\begin{adjustbox}{width=\linewidth}
\begin{tabular}{lrrrrrrr}
  Tool      & \mcrot{1}{l}{70}{Trivy} & \mcrot{1}{l}{70}{Grype} & \mcrot{1}{l}{70}{DepScan} & \mcrot{1}{l}{70}{Scout} & \mcrot{1}{l}{70}{Snyk} & \mcrot{1}{l}{70}{OSV} & \mcrot{1}{l}{70}{Vexy} \\
  \midrule
  Databases & 18 & 10 & 5 & 20 & 9 & 15 & 2 \\
  \bottomrule
\end{tabular}
\end{adjustbox}
\captionof{table}{Number of vulnerability databases per tool.}
\label{tab:number_of_databases}
\end{minipage}\hfill
\begin{minipage}[t]{.48\linewidth}
\centering\footnotesize
\begin{adjustbox}{width=\linewidth}
\begin{tabular}{lrrrrrrr}
  Tool   & \mcrot{1}{l}{70}{Trivy} & \mcrot{1}{l}{70}{Grype} & \mcrot{1}{l}{70}{DepScan} & \mcrot{1}{l}{70}{Scout} & \mcrot{1}{l}{70}{Snyk} & \mcrot{1}{l}{70}{OSV} & \mcrot{1}{l}{70}{Vexy} \\
  \midrule
  Trivy   & 1    &       &       &       &       &       &     \\
  Grype   & 0.12 & 1     &       &       &       &       &     \\
  DepScan & 0.11 & 0.13  & 1     &       &       &       &     \\
  Scout   & 0.28 & 0.18  & 0.08  & 1     &       &       &     \\
  Snyk    & 0.25 & 0.10  & 0.27  & 0.15  & 1     &       &     \\
  OSV     & 0.15 & 0.04  & 0.05  & 0.29  & 0.20  & 1     &     \\
  Vexy    & 0    & 0     & 0.16  & 0     & 0.10  & 0     & 1   \\
  \bottomrule
\end{tabular}
\end{adjustbox}
\captionof{table}{Jaccard index based on database coverage.}
\label{tab:database_similarity}
\end{minipage}
\vspace{-27pt}
\end{table}

\section{Discussion}

In this work, we have demonstrated a high level of inconsistency among reports from VEX-generation tools and examined several possible factors that could have helped explain this condition. 
The practical impact of inconsistent VEX-reports is significant for software supply chain security. Divergent results reduce trust in automated vulnerability management, making risk assessment and remediation decisions less reliable. Inconsistency across tools thus becomes an operational problem that weakens the usefulness of SBOM and VEX data in real-world security workflows, making understanding of the factors causing inconsistency an important vector in improving software supply chain security.

The first factor is the completeness and accuracy of the SBOM, which should list all components in the software. However, related work \cite{sbom_vuln,Balliu_2023,syft_sbom_chall,ozkan2024supplychaininsecuritylack,SBOM_python} highlight the challenges in creating complete and accurate SBOMs. In the work by Cofano et al. \cite{SBOM_python2} the effect of the SBOM precision on vulnerability identification for Python ecosystem was explored, leading to the conclusion that 80\% of vulnerabilities can be lost due to low SBOM accuracy. 

To explain the other factors, we should refer to the broader concept of the naming problem, mentioned in Section \ref{analysis:db}, which arises in several dimensions during the VEX-generation workflow, in addition to SBOM components. 
SBOM is affected by one dimension of the naming problem, related to different component identifiers. Different SBOM production tools apply various identifier categories, such as SWID tagging, PURL, or CPE \cite{SBOM_naming}. Thus, the same component (i.e., dependency) might be listed under different names in SBOM reports and considered as two different entities. This also leads to us not matching the reported vulnerabilities. However, in Section \ref{analysis:input-dep}, we show that this does not explain all the observed inconsistencies.

The next dimension of the naming problem appears in the VEX-generation stage during component mapping to the vulnerability databases. If the component has several identifiers, the vulnerabilities associated with each particular identifier may also vary depending on the name by which the component is known in the vulnerability database. 
The importance and general effect of SBOM-generation on vulnerability mapping was investigated in \cite{SBOM_Generators_impact}. In total, with the current precision of SBOM-generators only 20\% of vulnerabilities can be identified.

The naming problem also affects vulnerability identifiers. This problem was discussed in detail in Section \ref{analysis:impact-identifiers}. 
In our experiments, even though we observe an increase in similarity scores for reports based on a single vulnerability identifier system, the similarity scores are still low. This indicates the possible impact of other naming problem dimensions mentioned above.

Research focused on resolving naming inconsistencies in different layers could enhance vulnerability identification and VEX-report generation, leading to more consistent and reliable results.

Another key factor affecting the consistency of the reports is the reference to different vulnerability databases. We have found a relatively strong positive linear correlation (0.88) between similarity in vulnerability database references and similarity in reports. When tools map to the same vulnerability databases, we observe greater uniformity in the generated VEX-reports. Here we note that inconsistency is not, by definition, a bad thing. 

Vexy’s low similarities with other tools highlight its outlier behavior. Its reliance on limited databases and exclusive SBOM input likely restricts coverage. Future analysis and research could help clarify whether this discrepancy is due to narrow data sources, design interpretation, or strict validation rules.
We note that our study does not address the challenge of false positives and negatives in vulnerability identification. In fortunate circumstances, this could explain and resolve the inconsistencies we have found, but it could also contribute to the inconsistencies.

While SBOM quality showed to not be a major factor causing inconsistency, future work should examine its influence on VEX accuracy and completeness. Expanding the dataset with more tools and industrial cases, and studying how format harmonization (e.g. SPDX vs CycloneDX) affects consistency, would further strengthen these findings.

\section{Related Work}\label{related_works}

VEX-report studies are inherently linked to SBOMs, as VEX-reports disclose identified vulnerabilities in a project's components and dependencies, and SBOMs enumerate these components. 

Although the concept of SBOM specification is now widely recognized within both industry and academia, VEX remains relatively nascent. VEX is referenced as a component of SBOM in "The Minimum Elements For a Software Bill of Materials" published by the National Telecommunications and Information Administration (NTIA) \cite{ntia_sbom_min_req}. Consequently, several scholarly articles addressing SBOM-related issues reference VEX, often citing standards from NTIA or the Cybersecurity and Infrastructure Security Agency (CISA). 

Xia et al. \cite{xia2023empiricalstudysoftwarematerials} discern the perceptions of SBOM practitioners and the obstacles encountered in their implementation. Although not specifically addressing VEX itself, this paper illuminates the issue of SBOM generation, which is foundational to VEX quality. 
Eggers et al. \cite{osti_1901825} provide an exhaustive examination of SBOM applications within nuclear power facilities, mentioning VEX as part of vulnerability attestation. 
Yu et al. \cite{sbom_gen_metadata_dif_approach} examine the accuracy of SBOM generation solutions and briefly mention VEX as an SBOM add-on.
Dunlap et al. \cite{dunlap2023s3c2summit202302industry} express concerns regarding the inadequacy of current SBOM generation tools and the current state of VEX-reports generation. 
Raihanul \cite{sbom_analysis_supply_chain} analyzes the VEX specification and its formats, including its correlation with SBOMs. 
Williams et al.~\cite{research_directions} explore VEX as a prospective research trajectory, discussing its accuracy and trustworthiness. 
The importance and general effect of SBOM-generation on vulnerability mapping was investigated by Benedetti et al.~\cite{SBOM_Generators_impact}. In total, with the current precision of SBOM-generators only 20\% of vulnerabilities can be identified.


Javed and Toor \cite{javed2021understandingqualitycontainersecurity} explore the capabilities of Grype, Clair, and Dagda (these tools are vulnerability scanners for containers), focusing on their efficacy in detecting vulnerabilities. The authors calculated the amount and distribution of vulnerabilities found in 59 containers. However, their focus was primarily on Java-oriented packages.
In another publication \cite{10.1145/3481646.3481661}, the same authors evaluate vulnerability detection tools using the Detection Hit Ratio (DHR) metric. This metric assumes that the tool that identifies the greatest number of vulnerabilities is the most precise, whereas tools that omit certain vulnerabilities are considered less accurate. The deployment of the DHR metric is attributed to the absence of a reliable ground truth dataset available to the authors. 
Their findings revealed that Grype, despite achieving the highest DHR score, still lacks 34\% accuracy. 

The study \cite{assess_sec_risks_ssc_using_sbom} offers a comparative analysis of Trivy and Grype, highlighting discrepancies in the distribution of vulnerable packages and CVSS scoring results for these tools. 
Kalaiselvi et al.~\cite{10200828} present an assessment of the vulnerability scanning tools Grype and Trivy for containers and propose a joint integrated solution. The authors calculate the total number of vulnerabilities detected by each tool and the total number of unique vulnerabilities, where Grype performed better. Their analysis does not examine the consistency factor among the tools and instead focuses on Grype performance, rather than choosing a broader toolset.
O’Donoghue et al.~\cite{sbom_vuln} present their findings on vulnerability analysis for containers based on SBOM. Key findings after analysis of 2313 container images revealed significant variability in the number of vulnerabilities reported for different SBOM generation tools and SBOM formats. In contrast, we investigate the consistency among VEX-generation tools in identifying 
vulnerabilities.

Dann et al.~\cite{OSS-challanges} investigated the performance of vulnerability scanners on common development practices i.e. forking, patching, re-compiling, re-bundling and re-packaging. Although the evaluation was Java-oriented, it shows that of the 2,505 findings, only 34\% were true-positives. These findings further highlight the inconsistency problem in vulnerability scanners.

But in total, the field is relatively new and has the potential for future comprehensive research.

\section{Conclusion}
We analyze the state-of-the-art vulnerability scanning tools applied to Docker container images. Our toolset includes seven vulnerability scanners with VEX production capabilities: Trivy, Grype, DepScan, Docker Scout, OSV, Vexy, and Snyk. To assess consistency in reports of detected vulnerabilities, we apply Jaccard and Tversky similarity indices.  The main takeaways include:\\
\renewcommand{\labelenumi}{\textbf{\arabic{enumi}:}}
\begin{inparaenum}
    \item The results indicate a low level of pairwise similarity between the tools based on the Jaccard index.\\
    \item We observe the highest similarity scores between the VEX-reports of Trivy and Grype (69.4\%), while Vexy and Snyk have 0\% similarity.\\
    \item The highest similarity for vulnerabilities with CVE-only identifiers is 76\%, also between Trivy and Grype.\\
    \item The general similarity among all the tools is 0\% for our dataset.\\
    \item We have ruled out SBOM quality as an explanatory factor for the tools' inconsistency.
\end{inparaenum} 

Even if it may be expected that the tools produce slightly different results for various reasons, it is quite surprising that the similarity scores are so low. From a practical point of view, it is important to know that two tools that claim to do the same thing may produce largely disjoint results.
The most impactful issue affecting consistency in the results is the mapping and naming of vulnerabilities in different databases. However, limiting the dataset to a single vulnerability identifier only moderately increases similarity scores between tools. Our experiments highlight significant diversity in vulnerability detection, underscoring the immaturity of this product field. The main recommendation, which could be made based on the analysis, use several scanners with relatively low similarity scores to have a better coverage of the vulnerabilities.
%
%

\vspace{15pt}

\begingroup 
\let\clearpage\relax 
\bibliographystyle{unsrt}
\bibliography{bibliography} 
\endgroup

\end{document}